# Geography and the Internet:

# Is the Internet a Substitute or a Complement for Cities?

Todd Sinai
Joel Waldfogel
The Wharton School
University of Pennsylvania
and NBER

October 27, 2000

## ABSTRACT

By combining persons around the world into a single market, the Internet may serve as a *substitute* for urban agglomeration. That is, the Internet may level the consumption playing field between large, variety-laden and small, variety-starved markets. However, if local content on the Internet is more prevalent in larger markets, then the Internet may be a *complement* for urban agglomeration. Characterizing the nature of available content using Media Metrix web page visits by about 13,500 households, we document that substantially more online local content is available in larger markets. Combining this with CPS Internet use data, we find statistically significant direct evidence of both complementarity and substitutability: Individuals are more likely to connect in markets with more local online content; and holding local online content constant, are less likely to connect in larger markets. We also find that individuals connect to overcome local isolation: Blacks are more likely to connect, relative to whites, when they comprise a smaller fraction of local population, making the Internet is a substitute for agglomeration of preference minorities within cities, if not cities themselves. On balance we find that the substitution and complementarity effects offset each other so that the Internet does not promote or discourage agglomeration in larger markets.

We are grateful to the Electronic Commerce Forum at Wharton for giving us access to, and technical assistance with, the Media Metrix data. Sinai received financial support from the Ballard Scholars Program at the Zell-Lurie Real Estate Center, and Waldfogel thanks the Electronic Commerce Forum for financial assistance. Sam Chandan provided excellent research assistance.

Traditionally, markets for news and information as well as some retail goods have been predominantly local. As a result, consumers' welfare has been limited by the size of their local market, and agglomeration of persons sharing similar preferences has improved their welfare by facilitating the provision of products they want.[1] By agglomerating persons around the country – indeed, around the world – into a single market, the Internet offers the potential to radically alter consumption possibilities. In particular, the Internet may serve as a substitute for urban agglomeration by leveling the consumption playing field between large, variety-laden and small, variety-starved markets. But this is not necessary. Leveling the field requires that content on the Internet be similarly attractive to persons in large and small markets. If the Internet offers local, as well as general, information, then its role as a substitute for agglomeration will be undermined. Indeed, if local online content is sufficiently attractive – and if it is more prevalent in larger markets – then the Internet may be a *complement* for urban agglomeration.[2]

In this paper we ask whether the Internet serves as a substitute or a complement for urban agglomeration. Existing research indicates that larger markets provide greater, and smaller markets less, offline product variety. As a result, a consumers' welfare has traditionally been limited by the size of her local market. Moreover, if people have preferences unlike their neighbors', the availability of products they value has been limited not simply by the absolute size of their local markets, but rather by the size of the local constituency sharing their preferences.[3] In the past few years, consumers around the US have connected to the Internet in astounding numbers, at least in part for things the cannot obtain locally offline.[4] If individuals are more likely to connect as they live in smaller markets, then we can infer that the Internet is a substitute for cities. Furthermore, if persons isolated from others with similar tastes are more likely to connect, then we can infer not only that the Internet is a substitute for cities but also that it is a substitute for "product ghettoes," or groups of people within a market sharing in common their locally atypical product preferences.

---

[1] We are aware of other arguments for agglomeration besides consumption. Henderson (1974) presents a model in which city size balances production benefits against congestion costs. See Ciccone and Hall (1996) for recent empirical evidence on the production benefits of agglomeration as well as additional references.

[2] We are not the first to pose this question. See Kolko (1999), as well as Gaspar and Glaeser (1998), who find that telephones complement agglomeration because phone conversation complements face-to-face (two-way) communication. Our focus in this paper is on one-way communication over the web rather than two-way communication, but our question is similar.
1

At the same time that the Internet may substitute for cities by allowing access to distant content, a greater quantity of local content targeted at larger markets may nevertheless make the Internet a complement for cities. To explore this possibility, we ask a number of questions. First, how much Internet content is local? Second, is there more locally targeted content in larger markets? If so, then Internet content may complement urban agglomeration. This leads to our third question: does local content attract people to connect? The push of local offline product paucity – and the pull of local online variety – have opposing effects on whether the Internet is a substitute or a complement for cities. We separately document these competing effects and determine, on balance, which is stronger.

We first examine evidence about complementarity. Characterizing the nature of available content using Media Metrix data on 12 million web page visits by about 13,500 households in 138 local market areas in February 1999, we document that substantially more online local content is available in larger markets. Combining the Current Population Survey's 1998 Computer and Internet Use module with the Media Metrix-derived data on local online content, we document that local online content attracts people to connect. When we separately account for both local online content and our measure of local offline variety (population), we find statistically significant direct evidence of both complementarity and substitutability: Individuals are more likely to connect in markets with more local online content, and holding local online content constant, are less likely to connect in larger markets. On balance we find that these effects offset each other so that the Internet does not promote or discourage agglomeration in larger markets. The relationship between Internet connection and market size stands in stark contrast with evidence on local media markets, where the tendency to consume increases in the size of the market. We find no evidence that persons in larger markets are more likely to connect.

---

[3] For example, larger markets have more and better local newspapers (George and Waldfogel, 2000) and radio broadcasts (Waldfogel, 1999). These studies document that black and white consumers' welfare, in their capacity as media consumers, increase in the size of their own respective group populations.

We further explore whether the Internet allows individuals to overcome local isolation by asking how the tendency to connect varies across geography with the extent of individuals' local offline options, measured by the fraction of local persons of their race. We find that blacks are more likely to connect, relative to whites, as blacks are a smaller fraction of local population, suggesting that the Internet is a substitute for agglomeration of preference minorities within cities.

Friedman (1962) has argued that each person gets what she wants through market allocation, so that markets avoid the tyranny of the majority endemic to collective choice. Friedman's argument holds literally only when production can take place at arbitrarily small scale, so that available product variety does not depend on the size, or the preference composition of potential customers in the market. When fixed costs are sizable, the number of available products, and the resultant welfare of consumers in local markets can depend on the size and composition of the local market. By agglomerating consumers into larger markets, the Internet allows locally isolated persons to benefit from the product variety made available for consumers elsewhere. By increasing the size of markets relative to fixed costs, the Internet may therefore bring market allocation nearer to the ideal in which an individual's welfare does not depend on either the number of her neighbors or their product preferences.

---

[4] By December 1998, a third of individuals in the US had access to the Internet, either at home or elsewhere. See http://www.ntia.doc.gov/ntiahome/fttn99/contents.html .

The paper proceeds in six sections. Section 1 reviews available evidence on product variety and market size and characterizes the decision to use the Internet, as a function of one's preference type and the quality of local options. Section 2 describes the CPS and Media Metrix data used in this study. Section 3 employs the MM data to quantify local content on the Internet and, in particular, to characterize how the availability of local content varies with market size. Section 4 employs the Current Population Survey (CPS) data to characterize the demand for Internet connection. In particular, we examine how individuals' Internet connection tendencies vary with the extent of one's local and online options. A brief conclusion follows.

### I.  How Does the Internet Function as a Substitute or Complement for Cities?

*1. The Internet as a Substitute for Cities*

When production entails fixed costs and preferences differ across individuals, the number of differentiated product options available locally will increase in the size of the market.[5] Larger markets have more local product variety than small markets, and this greater variety draws a higher fraction of persons to consumption of local offline products. In this way persons benefit each other through what has elsewhere been termed a "preference externality" (Waldfogel, 1999).

By aggregating people in disparate locations into a single market, the Internet has the capacity to increase market size relative to fixed costs.[6] This can, in turn, raise the number of available products and reduce the dependence of consumer welfare on the number and mix of consumers in her local market. That is, consumers in small offline markets can instead turn to the Internet for products unavailable offline locally. What sorts of sites make the Internet a substitute for cities? We have in mind sites that offer content that is not geographically specific but which may have greater appeal in smaller markets with less offline product variety. For example, Spinner.com offers 140 channels of streaming music programming, over twice the number of radio stations available in any of the largest US markets. Spinner.com may appeal to listeners in both large and small markets but is presumably provides more of a benefit to listeners in small markets with few traditional radio stations. News sites, such as CNN.com or MSNBC.com, present domestic and international news of interest to individuals in cities of all sizes. But because small markets tend to have slender local newspapers (George and Waldfogel, 2000), people who live in them may place a higher value on the availability of news on the Internet.

---

[5] This is what Spence (1976a,b) terms the "product selection problem." See also Dixit and Stiglitz (1977). Waldfogel (1999) presents an empirical characterization of how product availability varies with the size and demographic composition of local markets.

[6] Computer technology may also reduce the absolute size of at least the exogenous component of the fixed and sunk costs of operating a business. Given the large advertising expenditures of web retailers such as Amazon.com, it is not clear whether web businesses have lower fixed costs than bricks and mortar businesses, when endogenous fixed costs are taken into account. See Sutton (1991) for extensive discussion of endogenous fixed costs.

*2. The Internet as a Substitute for Product Ghettoes*

The paucity of offline product variety is not determined solely by the total population in an area. To the extent that preferences differ across types of individuals, the number of like-minded persons in a local area will determine the size of the offline market and the amount of locally available offline products that would appeal to those people. Since the distribution of types differs across geographic markets, we expect persons to be more likely to connect to the Internet to satisfy their locally unfulfilled tastes when they are "preference minorities," that is, part of a group with distinct preferences that makes up a small number of the local population. For example, it is well documented that blacks and whites have sharply different preferences in some categories of products. In particular, the radio programming formats attracting two thirds of black listening collectively attract less than 2 percent of nonblack listening. In major cities with both tabloid and non-tabloid newspaper options, the tabloid attracts about three quarters of black readers, compared with about a third of nonblack readers.[7] In these contexts, the satisfaction of blacks as consumers of local products has typically depended on the number or share of blacks in the local population, not the overall size of the local market.[8]

*3. The Internet as a Complement for Cities*

In addition to providing universal content that appeals to individuals in any size market, the Internet may also be a local medium. The Internet can provide information that helps people to navigate cities, and may deliver other goods and services that improve city life. For example, city portals, such as boston.citysearch.com, provide information about events, restaurants, and movie listings. Match-making sites, such as boston.matchmaker.com, help users in large cities meet people. And sites associated with local newspapers or television stations provide another distribution channel for local news. If there are fixed costs associated with producing such content, then the number and variety of local sites may increase in the size of the local market, making the Internet more useful to people in larger markets and mitigating the Internet's role as a substitute for local offline product variety.

These considerations motivate the four questions that this study addresses. How much web content is local? Is there more local online content in larger markets? How does the tendency to connect vary with one's local offline and online options? Which effect dominates?

---

[7] Based on views about public spending blacks and whites have substantially different political preferences as well (Oberholzer Gee and Waldfogel, 2000).
[8] With some products the absolute number of local persons determines the number of available options. This is true for radio broadcasting. See Berry and Waldfogel (1999ab) and Waldfogel (1999). For other products, such as daily newspapers, fixed costs rise endogenously with market size, keeping the number of products small. George and Waldfogel (2000) present evidence that fraction black in the market determines the how black-targeted the paper(s) is (are). Other statistics in this paragraph are also drawn from these sources.

**II. Data**

Data for this study are drawn from two sources, the 1998 Current Population Survey Computer and Internet Use Supplement and a February 1999 Media Metrix data extract. The CPS supplement has information on Internet connection, as well as demographic and geographic information, for 123,000 individuals in December 1998. For survey respondents that have Internet connections, the CPS reports whether they regularly use the Internet for nine categories of activities, including shopping; news, information, and weather; and school courses. From the nearly 123,000 individuals in the CPS data, we reduce our sample to 86,523 by restricting our attention to those households that live in metropolitan statistical areas (MSAs) that can be matched to designated market areas (DMAs) in our Media Metrix data set.

Table 1 reports sample characteristics. Because we are interested in factors that may affect the likelihood of a household using the Internet, we first define a person as being Internet connected if she has access at home. Almost 32 percent of the persons in our sample have such connections. Just over half the sample has one or more computers at home. Almost 13 percent of the sample is black, and 24 percent is college-educated.

The third of the sample that is Internet-connected is disproportionately white and is more highly educated than the population as a whole.[9] Only about 5 percent of the connected sample is black, and approximately 37 percent is college educated. Those persons who have Internet connections at home are slightly less likely to be female than the overall population. Naturally, nearly all of the connected sample has a home computer.

---

[9] These results are consistent with evidence elsewhere on the digital divide. See http://www.ntia.doc.gov/ntiahome/fttn99/contents.html .

The bottom part of table 1 breaks down the fraction of connected persons who report regularly using the Internet for each of nine activities: email, courses, news and weather, phone, search, job search, job tasks, shopping and paying bills, and "other." The most widespread activities are email (55 percent), search (42), and news and weather (32). The least common activities are phone (4) and job search (10).

Our second data source follows the Internet usage behavior of a panel of households. Media Metrix collects data on all web page visits by a representative sample of households in 211 designated market areas (DMAs) by placing recording software on panelists' computers. In our extract, which covers February 1999, each visit to a web page by a household is a separate record – with over 12 million page visits in total during the month.[10] Media Metrix reports basic information about their Internet-connected panelists, such as DMA, income category, educational attainment, and race, to their data on web surfing. Table 1 reports the education distribution for the Media Metrix sample, and it is similar to the CPS sample of households with internet connections at home.[11]

In addition, for each site visit, we observe the URL, or "address" of the web page, which Media Metrix classifies into one of 13 categories.[12] The first column of table 2 reports the distribution across these categories of page hits and numbers of visited sites, omitting the "other" category as we exclude it from our sample during the estimation. The category with the most hits is news, information and entertainment, with 21.7 percent of the 11,809,482 total hits, followed by search engines (14.6), marketing/corporate (12.5), shopping (12.4), web service providers (10.9) and adult content (10.5). Directory sites have the fewest hits with only 0.9 percent of the total.

---

[10] A "site," in our extract, is typically a three-level name, such as www.aol.com. The data contain other sites at America On-line (AOL), such as members.aol.com, as separate "sites," even though they are in the same "domain."
[11] Media Metrix also reports race data for 5736 of the 13,509 households with valid data on all fields required in this study. Of households reporting race, 2.4 percent are black.
[12] The Media Metrix categories are: web service providers, commercial online networks, search engines, government, education, adult, marketing/corporate, news/information/entertainment, shopping, other, travel/tourism, internet service providers, and directories.

Turning to the number of sites visited in each category, reported in column 2, "news, information, and entertainment" has the largest share of sites with 20.4 percent of the 52,282 total sites the Media Metrix panelists visit. However, search engines and shopping sites, which received large portions of the total number of page hits, comprise a very small fraction of the sites in the sample, 2.5 and 3.8 percent respectively. On the other hand, education and government sites, which each attract around 2 percent of the web traffic, account for 18.9 and 5.6 percent of all sites visited by the panelists. This pattern suggests that there may be more concentration in shopping and search engines, with a few sites each receiving a large amount of traffic. Government and education sites seem to appeal to a more fragmented audience, with many sites each having a small number of visits.

### III.     How Much Content is Local?

To determine whether the Internet is a complement to cities, we need to measure the amount of local content targeted at each metropolitan area. As our metric, we count the number of sites that produce content that appeals primarily to one particular market. Unfortunately, there exists no comprehensive list of sites by locale from which one could characterize local content.[13] Indeed, one cannot determine the localness of a site's targeting from the registration location of a site, or where the parent company's headquarters are located, since the site's visitors could be from anywhere.

---

[13] Kolko (1999) uses the list of registered domain names and shows higher "domain density" in larger markets, which is at least suggestive that web content is complementary with cities. Domain registration indicates the geographic location of the registrant, not the site users, however.

Fortunately, we can use the Media Metrix data to measure the geographic focus of a site. By recognizing that a locally targeted site must have a primarily local audience, we can use the geographic origin of a site's visitors as reported in the Media Metrix data to estimate the extent of its local focus and which market it primarily serves. In essence, after calculating every site's share of each market's total page hits, we presume that a site that has a sufficiently high proportion of its total market shares across all markets coming from just one market must be providing content of local interest to that market.

To make this more concrete, we compute each site's index of "site localness" as

$$1 \bigg/ \sum_j \left[ \frac{p_{ij}/p_j}{\sum_j (p_{ij}/p_j)} \right],$$

where $p_{ij}$ is page visits to site $i$ from DMA $j$, $p_i$ is the total page visits (by households in any DMA) to site $i$, and $p_j$ is the total visits (to any site) from DMA $j$. This formula is essentially an inverse HHI scaled to reflect the fact that population varies across DMAs. If we did not normalize by each DMA's total page visits, sites would appear to be targeted towards larger DMAs even if they truly were equally appealing to any individual in any part of the country. When all visitors are from a single DMA, the index is 1. If a site has equal market shares in two DMAs, the index is 2. The larger the index, the less local a site. A perfectly geographically neutral site's index equals the number of DMAs in the sample, 138. We then attribute a local site, which we define as having a "localness index" of two or less, to the locale that contributes the site's largest market share.[14]

Because our measure of localness depends on the composition of a site's audience, our accuracy in classifying sites as local diminishes for sites with very few visits from the Media

---

[14] The most obvious type of local site that we misclassify as not local are sites which contain local information for a number of locales. Since we require that a site be targeted to one locale to be defined as local, these sites do not qualify. However, we suspect the narrowly targeted sites are a reasonable proxy for the localness patterns we would find if we used a more broad definition.

Metrix panel during the month. To counteract that problem, we require that sites exhibit a minimum level of interest, as measured by page hits, in the DMA that they target in order to be considered local to that market. Since Media Metrix does not sample proportionally to market population, the cutoff increases in sampling frequency. For example, we require that sites that are considered to be local to New York city to have at least 100 hits from New Yorkers during the month. For Boston, local sites must have at least 100 hits times the ratio of Boston's sampling frequency (Media Metrix households divided by the DMA population) to New York's sampling frequency. This way, even if Boston has fewer households than New York in the Media Metrix panel, our threshold maintains the same economic importance across locations.

When we look at the sites with the largest and smallest adjusted localness indices across Media Metrix categories, the index produces reasonable results. The least local sites have localness indices over 50 and include such familiar sites as hotmail.com (web service provider), aol.com (commercial online network), yahoo.com (search engine), irs.ustreas.gov (government), microsoft.com (marketing/corporate), msnbc.com (news, information, entertainment), ebay.com (shopping), expedia.msn.com (travel), worldnet.att.net (ISP), and infospace.com (directory).[15]

The most local sites have localness indices close to one and tend to refer to locales in their URL. For example, chicago.email.net (web service provider), det.state.vt.us (government), mail.business.pitt.edu (education), community.oregonlive.com (news/information/entertainment), and az.rmci.net (ISP) all have localness indices equal to one meaning that all their hits came from one DMA. Many other local sites are associated with local newspapers or television stations.

The sixth and ninth columns of table 2 show the distributions of hits to local sites, and local sites themselves, across Media Metrix categories. News, information and entertainment has the largest number of local hits and sites. Internet service providers, which often serve a very small area, are next in terms of sites but not in hits.[16] Education and web service providers are the next most local. Commercial on line networks and directory sites are the least local, with almost no sites considered to be locally targeted. On average, 4.3 percent of all hits are to the 985 sites that we deem local. The education category has the highest share of its site visits going to local sites while news, information, and entertainment, which has the highest share of all local hits, has only 6.4 percent of its hits going to local sites.

The average number of local sites per DMA, reported in table 3, follows the same general pattern. Of the 6.12 local sites on average in each DMA, news, information, and entertainment sites account for 1.76, education for 0.99, and ISPs for 0.97. The DMA with the most local sites, New York, has 39. No DMA has more than 12 local news, information, and entertainment sites or 12 education sites.[17]

We finish our description of the local site data by asking how the amount of local content varies with the size of the local market. If there are fixed costs of providing a local online site, then the quantity of local online content will increase in the size of the local connected market. In traditional media, as we have mentioned above, larger markets have more local content (more radio stations, more and better local newspapers. Are there similarly more local online sites in larger markets? This question is interesting both in its own right and because we seek an instrument for local online content in subsequent analyses of its effect on the tendency for users to connect to the Internet.

---

[15] We have suppressed the "www" site name prefix for clarity.
[16] ISPs have slightly fewer local hits than do adult sites. We have not visited the adult sites to verify their local content.
[17] Table 3 reflects the 138 DMAs in our final sample. Our sample was constrained because the CPS reports data by MSA and Media Metrix by DMA. Using an MSA-Zip Code-DMA correspondence, we were able to successfully map 138 MSAs to the DMA that contained them. In the cases where an MSA mapped to more than one DMA, we allocated the DMA with the largest share of the MSA population to the MSA. Our population figures are aggregated up from the Zip Code level by DMA.

Table 4 reports regressions of DMA local online content, overall and by category, on total DMA population. An additional million residents in a DMA adds more than two local sites. While the relationship between DMA population and the number of local sites is positive and significant for most of the Media Metrix categories, the size of the effect is largest for news/information/entertainment. It is also sizable for education sites, adult sites, ISP sites, and web service provider sites.[18] That larger markets have more Internet content indicates that the Internet is not simply a leveler of utility across geography and, indeed, may be a city complement.

IV. **The Demand for Internet Connection**

If the Internet is a substitute for cities, the probability that a person connects to the Internet should increase as the variety or quality of local offline options decline. If it is a complement, it should be more prevalent in larger cities and people should be more likely to use it where there are more city-specific options. This section asks how the tendency to connect to the Internet varies with measures of the quality of local online and offline options.[19] Our basic measures of the extent of offline options are total local population, which is presumed to increase the variety of goods and services available. In addition, a resident's relevant product variety is determined by the size of the market of people who share her preferences. In our estimation, we will measure that market with the population – and population share – of one's group, where the groups are blacks and nonblacks.

---

[18] The magnitude of the estimated coefficient appears be related to our priors on how well our localness index captures the number of local sites in the category. In categories in which we think the index does fairly well, such as news/information/entertainment, education, and ISPs, the result is fairly strong. In categories where one would be surprised to find any locally-targeted sites, such as search engines, travel/tourism, and directories, our estimated coefficient is basically zero.

[19] One might in principle study demand for Internet connection as a function of price or availability of ISPs. Greenstein (1999) indicates that by 1998 Internet access is widely available in all MSAs. The price of Internet access also varies little across MSAs. A regression of the 1998 CPS measure of monthly ISP costs (hesiu9) on 1990 MSA population gives a constant of $17.46 (s.e.=$0.21) and a population coefficient indicating that the price paid for access increases by $0.043 ($0.018) per million of additional population.

Our measure of local online product variety is the number of local sites. Although the first panel of table 4 shows that the Internet provides more local content in bigger places, it does not say whether the Internet actually enhances city life. For that to be true, people must want local content. There is ample evidence in traditional local media that the greater variety brought forth in larger markets attracts a higher fraction of the population to consumption. The radio listening, and newspaper reading, shares are higher in larger markets. The greater quality and variety of options in traditional media provide part of the reason why persons' welfare, in their capacity as media consumers, is higher in larger markets. Thus we turn to whether this pattern arises with Internet content: does the greater variety of online options targeted at big-city consumers attract a higher fraction of them to the Internet? If so, then the Internet functions as a city complement.

*1. Internet Connection and the Extent of Local Offline and Online Options*

We examine how the tendency to connect is affected by local offline variety by using the CPS data to estimate a linear probability model of an individual having a home Internet connection as a function of DMA population:

$$C_i = \alpha + \beta \cdot POP_m + \delta \cdot X_i + \varepsilon_{im} \qquad (1)$$

The left-hand-side variable, $C$, takes the value of one if individual $i$ has an Internet connection at home. Our basic measure of local offline product availability, DMA population, is denoted by *POP* and varies only at the DMA level, $m$. In some specifications, we add a large set of individual level demographic controls, $X_i$, including race, gender, income dummies, education dummies, age dummies, dummies for the presence and age of children, occupation and industry dummies, and variables reflecting household composition. In those specifications, the estimated coefficient on *POP* will measure the effect of offline variety in the local market on individuals' decisions to connect to the Internet even after accounting for their own characteristics.

The first column of table 5, which includes only DMA population as a covariate, shows that overall the probability of connecting to the Internet does not vary with population, as the point estimate is slightly negative but indistinguishable from zero.[20] This substitution and complementarity effects of the Internet are either nonexistent or offsetting. Supplementing this specification with individual controls in column (5) has little substantive effect. The coefficient on MSA population in column (5) is slightly more negative and roughly 1.25 times its standard error.

We distinguish the substitute and complement effects by adding a measure of the number of local online options. Columns (2) and (6) add the number of local news, information, and entertainment sites ($LOCAL_m$) in the DMA to the specifications in columns (1) and (5), resulting in the following equation:

$$C_i = \alpha + \beta \cdot POP_m + \gamma \cdot LOCAL_m + \delta \cdot X_i + \varepsilon_{im} \tag{2}$$

In this regression, β measures the effect of substituting towards the Internet when there is less local variety and γ reflects the value of local content as measured by the likelihood of connecting to the Internet. In both the specifications in columns (2) and (6) the coefficient on local population is negative and significant, indicating that the Internet functions as a substitute for cities, although the magnitude of both the substitution and complement effects are smaller in column (6). The coefficient on local online sites is positive and significant in column (2), indicating that local online content attracts people to the Internet. Thus our finding of a zero net effect of population on Internet connection is due to the substitute and complement effects offsetting each other.

---

[20] The standard errors in all of these and subsequent regressions are adjusted for clustering on DMAs.

When we add the individual level controls in column (6), we still obtain a positive coefficient but our t-statistic declines to 1.61. Although not statistically significant at a 95 percent confidence level, we consider this result – which is remarkably stable given that we added a tremendous number of individual-level controls – to be quite convincing that there is a complement effect. It should be emphasized that the apparent sample size of 75,311 observations considerably overstates the available variation in the data. In fact, the total number of local sites only varies across our 138 DMAs, and then only between zero and 12. We correct the standard errors for the fact that we have only DMA-level variation in our key explanatory variables which is why the statistical significance is weaker when the number of covariates increases.

Local online content is a potentially endogenous variable, because the supply of local sites may reflect a market response in areas where people are more likely to connect for some unobservable reason other than local content. Thus we need to instrument for $LOCAL_m$ with an exogenous, DMA-level variable. Any average characteristic of the DMA that affects the overall market for local content but does not have an independent effect on an *individual's* decision to connect (controlling for her observable characteristics) is a valid instrument. We will not be able to use total population as an instrument since it is our indirect measure of the quality of local offline options and provides no additional leverage for testing the relative effects of online and offline options on connection. We cannot use the size of the local connected population as an instrument for local online content because the connection decision may itself reflect the amount of local online content. However, we can use a measure of the population of persons likely to connect, those with four or more years of college.[21] [22] Our identification thus relies on the fact that the number of highly educated people in the DMA should not affect any given resident's decision to connect to the Internet except through the possibility that a large local potential Internet audience overcomes sites developers' fixed cost of providing more local web sites. This identification strategy is made stronger by controlling for the observable characteristics of the individual, so the instrument relies on the number of highly educated people in the DMA affecting the number of local sites but not independently affecting the individual's decision to connect conditional on her own level of education (and income, etc). We use the number rather than the share of highly educated people because the absolute audience size determines whether a site covers its fixed costs. A small DMA that is 50 percent highly educated may not have enough bodies to support a local web site when a large DMA that is only 10 percent highly educated would.

---

[21] According to the 1998 CPS, 46 percent of highly educated persons have Internet connection at home, compared with 25 percent of the remaining population.

Indeed, this instrument predicts the number of local sites well. Columns (3) and (7) report the first stages of an IV regression: the number of local news, information, and entertainment sites on DMA population and the number of highly educated persons, with column (7) including the full set of other explanatory variables as well. In both regressions, the number of local sites declines with total population but increases substantially with the number of highly educated people.

In columns (4) and (8), we report IV estimates of regressions of Internet connection on population and (instrumented) local online content, with and without the individual controls. As in the OLS regressions in columns (2) and (5), both these regressions show negative effects of population and positive effects of local online content, although the point estimates double in magnitude. However, while the results in column (4) are statistically different from zero, when we add the full set of covariates the t-statistic on the substitution effect declines to 1.51 and the complementarity effect is significant only at the 90 percent level. Still, the economic significance of the results is qualitatively the same as in column (4).

---

[22] It is possible that the highly educated population also increases local offline options. If so, this will tend to mitigate its effect on local online options and therefore bias the instrument against showing an effect of instrumented online content on connection

These results show that local content, which is more plentiful in larger markets, attracts people to the Internet. Holding constant the amount of local online content, people are less likely to connect as their local offline options, proxied by population, are more appealing. The Internet functions as both as a substitute and a complement for cities. On balance, as columns (1) and (5) show, the substitution seems to just slightly more than offset the complementarity.[23]

*2. Internet Connection and Racial Isolation*

While the Internet does not function, on balance, as a substitute for cities generally, it may still allow locally isolated individuals to surmount the limitations of their local offline markets. To put this another way, the Internet may be a substitute for "product ghettoes." To investigate this we ask whether racially isolated individuals are more likely to connect to the Internet. We implement this in table 6 by asking whether blacks (nonblacks) are less (more) likely to connect as blacks face less appealing local offline product options. Depending on the nature of the local product, its appeal to blacks might reflect either the proportion or absolute number of blacks in the local area. For products with large fixed costs relative to market size, the market supplies few options, and positioning relative to black preferences depends on the fraction black in the local market.[24] For products with smaller fixed costs relative to market size, the market can supply multiple options, and the appeal of the local offline product options depends on the absolute number of blacks.[25]

---

[23] When we use the total number of local sites across all categories as an explanatory variable we do not find a statistically significant result. However, this appears to be an artifact of some of the categories of local sites being poorly measured. When we restrict our attention to those categories with a sufficient number of obviously local sites, such as education and ISPs, consistent results emerge.
[24] George and Waldfogel (2000) document this as the mechanism for local daily newspapers.
[25] Waldfogel (1999) documents that the number of black-targeted radio stations, as well as the black tendency to listen to the radio, vary with the size of the local black population.

Because the content offered over the Internet may take either form, we perform tests of whether blacks use the Internet to overcome racial isolation using both black population percents and absolute levels.

First, we perform the test in proportions, regressing the connection dummy on the MSA black share, separately for blacks and nonblacks. We also run the test in absolute numbers, substituting the black and nonblack market populations for the black share. Columns (1) and (2), and (4) and (5), of table 6 report separate black and nonblack regressions of the tendency to connect on either the fraction black or the absolute numbers of blacks and nonblacks in the MSA. All of the regressions in table 6 include the full set of controls. None of these regressions give statistical significance, although blacks tend to be less likely to connect when they are absolutely or proportionately more numerous.

It is possible, however, that local market-level unobserved factors, such as local offline traffic congestion, affect the attractiveness of obtaining things locally. By extension, these factors would also affect the tendency for persons in that market to connect to the Internet. We can accommodate this by including a local market fixed effect. When we do this, we cannot identify effects of market-level factors, such as population, or the percent black in the market. We can, however, identify the *difference* between the effect of the MSA black share or the MSA's populations of blacks and nonblacks on black and white tendencies to connect as the coefficient on a black dummy interacted with, say, the MSA percent black.

Columns (3) and (6) of table 6 report estimates of the MSA fixed effects regressions, and we find a systematically negative coefficient on the interaction of the black dummy with the black population share. That is, relative to nonblacks, blacks are less likely to connect to the Internet as the black population share in the local market increases. We find similar results in absolute levels of population. These results provide strong evidence that persons are more likely to connect, the more they are isolated locally. These results, like the negative overall effect of population on the tendency to connect, indicate that connection is a substitute for living among persons sharing similar product preferences. This result suggests that the Internet is a substitute for product ghettoes.

*3. Computer Ownership, Market Size, and Racial Isolation*

Connection to the Internet requires two separate but presumably related decisions. First, one must purchase a computer or Web TV.[26] Second, given PC ownership, one must purchase Internet connection. If the decision to own a home computer were orthogonal to the Internet connection decision, it would make sense to run the regressions in the previous sections only on the sample of persons with a home computer. Instead, we suspect that these are related decisions (one reason to purchase a computer is to connect to the Internet). In table 7 we explore how the decision to own a home computer varies with our measures of the attractiveness of local offline and online options. Columns (1) and (4) show that computer ownership declines, although not significantly, in market size. Column (2) shows that computer ownership is higher in markets with more local online content and, conditional on local online content, lower in larger markets. In column (5), which adds controls to the column (2) specification, signs are the same, but the magnitude of the estimated coefficients declines, and they are not statistically significant. Columns (3) and (6) revisit the relationship between computer ownership and the determinants of online and offline options, now instrumenting local online content as above. In both cases the IV results reinforce the OLS results: results without controls are significant; results with controls have the same sign but are insignificant. Taken together, the results in table 7 indicate that part of Internet's relationship with city size operates through computer ownership.

---

[26] According to the 1998 CPS, 0.70 percent of the non-computer-owning population has Web TV.

Table 8 examines how computer ownership varies with racial isolation. The higher the black population share, the lower the tendency for blacks to own home computers, both absolutely and relative to whites. Columns (1) and (2) show that both blacks and whites are less likely to connect in markets with larger black shares. We interpret the white effect, as well as part of the black effect, as a reflection of unobserved heterogeneity. The specification in column (3) includes a market fixed effect and identifies the effect of interest as the relationship between the black-white deviation in computer ownership rates and the market's percent black. Results in absolute population numbers provide a check on the percent specifications. Estimates in columns (4)-(6), are similar to results in the first half. Therefore the role of the Internet in allowing persons to overcome preference isolation operates partly through the decision to own a computer.

## V. Conclusion

It is apparent from our results that, in spite of more and better local online options in larger markets, the tendency to connect to the Internet is not affected by market size. This result stands in sharp contrast to relationships in traditional media, which reinforce the consumption welfare advantages of larger markets. In the case of the Internet, local content does encourage increased connection in larger markets, as with traditional media. However, unlike traditional media, the Internet also provides access to a national level of variety for small places, mitigating the advantage of larger markets over smaller ones. This kind of effect is especially clear in blacks' tendency to use the Internet relative to whites to overcome preference isolation.

On balance, the Internet serves as a substitute for cities in general and product ghettoes in particular. Still, the quantity of local Internet content is larger in larger markets, which tends to make the Internet a complement for cities. These two effects almost completely offset each other, leaving individuals no more likely to connect to the Internet if they live in a smaller location. To paraphrase Mark Twain, reports of the death of cities at the hands of the Internet may be greatly exaggerated.

# References

Berry, Steven T. and Joel Waldfogel. "Free Entry and Social Inefficiency in Radio Broadcasting." *RAND Journal of Economics*. 1999.

Berry, Steven T. and Joel Waldfogel. "Public Radio in the United States: Does it Correct Market Failure or Cannibalize Commercial Stations?" *Journal of Public Economics*, 1999.

Borenstein, Severin. "On the Efficiency of Competitive Markets for Operating Licenses." *Quarterly Journal of Economics*. Vol. 103 (2). P. 357-385. May 1988.

Bresnahan, Timothy F; Reiss, Peter C. "Entry and Competition in Concentrated Markets." *Journal of Political Economy*. Vol. 99 (5). p 977-1009. October 1991.

Ciccone, Antonio and Robert E. Hall. "Productivity and the Density of Economic Activity." *American Economic Review*. Vol. 86 (1). p 54-70. March 1996.

Dixit, Avinash K; Stiglitz, Joseph E. "Monopolistic Competition and Optimum Product Diversity." *American Economic Review*. Vol. 67 (3). p 297-308. June 1977.

Friedman, Milton. *Capitalism and Freedom*. Chicago: University of Chicago Press, 1962.

Gaspar, Jess and Edward L. Glaeser. "Information Technology and the Future of Cities." *Journal of Urban Economics*. Vol. 43 (1). p 136-56. January 1998.

George, Lisa and Joel Waldfogel. "Who Benefits Whom in Daily Newspaper Markets?" Mimeo. University of Pennsylvania, August 2000.

Greenstein, Shane. "Building and Delivering the Virtual World: Commercializing Services for Internet Access." Mimeo. Kellogg School. 2000.

Henderson, J. V. "Optimum City Size: The External Diseconomy Question." *Journal of Political Economy*. Vol. 82 (2). p 373-88. Part I, March-April 1974

Kolko, Jed. "The Death of Cities? The Death of Distance? Evidence from the Geography of Commercial Internet Usage." Mimeo. Harvard University. 1999

Krugman, Paul. "Scale Economies, Product Differentiation, and the Pattern of Trade." *American Economic Review*. Vol. 70 (5). p 950-59. Dec. 1980.

Krugman, Paul. "Increasing Returns and Economic Geography." *Journal of Political Economy*. Vol. 99 (3). p 483-99. June 1991

Oberholzer Gee, Felix and Joel Waldfogel. "Tiebout Acceleration." Mimeo. University of Pennsylvania. August 2000.

Rogers, R. P. and J. R. Woodbury. "Market Structure, Program Diversity, and Radio Audience Size." *Contemporary Economic Policy* 14, no.1 (1996):81-91.

Siegelman Peter and Joel Waldfogel. "Race and Radio: Preference Externalities, Minority Ownership, and the Provision of Programming to Minorities." Mimeo. University of Pennsylvania. 1998.

Spence, Michael. "Product Selection, Fixed Costs, and Monopolistic Competition." *The Review of Economic Studies*. Vol. 43 (2). p 217-35. June 1976.

Spence, Michael. "Product Differentiation and Welfare." *American Economic Review*. Vol. 66 (2). p 407-14. May 1976.

Sterngold, James. "A Racial Divide Widens on Network TV." *New York Times*. December 29, 1998. P. A1.

Sterngold, James. "For All the TV Pilots, There's Just Not Enough Youth to Go Around." *The New York Times*, May 17, 1999. Page E1.

Sutton, John. *Sunk costs and market structure: Price competition, advertising, and the evolution of concentration.*. Cambridge, Mass. and London: MIT Press. 1991.

Tiebout, Charles. "A Pure Theory of Local Expenditures." *The Journal of Political Economy*. V. 64(5) (Oct 1956) pp. 416-424.

Waldfogel, Joel. "Preference Externalities: An Empirical Study of Who Benefits Whom in Differentiated Product Markets." NBER Working paper 7391, October 1999.

**Table 1: CPS and Media Metrix Sample Characteristics**

|  | CPS | | MM |
|---|---|---|---|
|  | all persons | connected persons |  |
| Internet at Home | 31.95% | 100.00% |  |
| Computer(s) at Home | 51.07% | 99.25% |  |
| One Computer at Home | 37.23% | 64.80% |  |
| Two Computers at Home | 9.88% | 23.78% |  |
| Three+ Computers at Home | 3.96% | 10.67% |  |
|  |  |  |  |
| White | 71.80% | 85.13% |  |
| Black | 12.01% | 4.76% |  |
| Native American | 0.63% | 0.55% |  |
| Asian | 3.95% | 4.89% |  |
| Hispanic | 11.61% | 4.66% |  |
|  |  |  |  |
| No High School | 38.15% | 32.03% | 31.2% |
| High School | 23.25% | 15.39% | 10.8% |
| Some College | 14.43% | 16.00% | 20.2% |
| College | 17.79% | 25.51% | 23.2% |
| Post Graduate | 6.38% | 11.07% | 14.5% |
|  |  |  |  |
| Female | 51.98% | 49.63% | 40.4% |
|  |  |  |  |
| Regular Use of Internet for: |  |  |  |
| Email |  | 54.98% |  |
| Courses |  | 25.39% |  |
| News, Weather |  | 31.99% |  |
| Phone |  | 4.24% |  |
| Search |  | 42.37% |  |
| Job Search |  | 10.09% |  |
| Job Tasks |  | 20.95% |  |
| Shop, Pay Bills |  | 17.81% |  |
| Other |  | 7.91% |  |
|  |  |  |  |
| Number of Obs. | 86,523 | 24,642 | 13,509 |

**Table 2: Distribution of Hits and Sites, by Category**

| | All Sites | | | | Local Sites | | | | | |
|---|---|---|---|---|---|---|---|---|---|---|
| | Hits | Pct of total | Sites | Pct of total | Hits | Pct of total | Pct of cat hits | Sites | Pct of total | Pct of cat sites |
| Web Service Provider | 1,283,586 | 10.9% | 5,524 | 10.6% | 50,484 | 10.0% | 3.9% | 85 | 8.6% | 1.5% |
| Commercial Online Network | 743,877 | 6.3% | 442 | 0.8% | 2,067 | 0.4% | 0.3% | 2 | 0.2% | 0.5% |
| Search Engine | 1,729,169 | 14.6% | 1,329 | 2.5% | 6,606 | 1.3% | 0.4% | 20 | 2.0% | 1.5% |
| Government | 224,440 | 1.9% | 2,940 | 5.6% | 8,058 | 1.6% | 3.6% | 29 | 2.9% | 1.0% |
| Education | 274,214 | 2.3% | 9,896 | 18.9% | 53,030 | 10.5% | 19.3% | 154 | 15.6% | 1.6% |
| Adult | 1,237,394 | 10.5% | 6,640 | 12.7% | 83,877 | 16.6% | 6.8% | 132 | 13.4% | 2.0% |
| Marketing/Corporate | 1,471,437 | 12.5% | 6,217 | 11.9% | 40,101 | 8.0% | 2.7% | 95 | 9.6% | 1.5% |
| News/Information/Entertainment | 2,567,529 | 21.7% | 10,678 | 20.4% | 164,924 | 32.7% | 6.4% | 274 | 27.8% | 2.6% |
| Shopping | 1,465,054 | 12.4% | 1,999 | 3.8% | 15,370 | 3.0% | 1.0% | 23 | 2.3% | 1.2% |
| Travel/Tourism | 177,925 | 1.5% | 714 | 1.4% | 1,961 | 0.4% | 1.1% | 7 | 0.7% | 1.0% |
| ISP | 533,632 | 4.5% | 5,103 | 9.8% | 77,398 | 15.4% | 14.5% | 163 | 16.5% | 3.2% |
| Directory | 101,225 | 0.9% | 800 | 1.5% | 294 | 0.1% | 0.3% | 1 | 0.1% | 0.1% |
| Total: | 11,809,482 | 100.0% | 52,282 | 100.0% | 504,170 | 100.0% | 4.3% | 985 | 100.0% | 1.9% |

**Table 3: Average Number of Local Sites, by Category**

| Category | Mean | Minimum | Maximum |
|---|---|---|---|
| Total Local Sites | 6.12 | 0 | 39 |
| Local Web Service Provider Sites" | 0.48 | 0 | 5 |
| Local Commercial Online Network Sites | 0.01 | 0 | 1 |
| Local Search Engine Sites | 0.12 | 0 | 2 |
| Local Government Sites | 0.20 | 0 | 2 |
| Local Education Sites | 0.99 | 0 | 12 |
| Local Adult Sites | 0.81 | 0 | 8 |
| Local Marketing/Corporate Sites | 0.58 | 0 | 6 |
| Local News/Information/Entertainment Sites | 1.76 | 0 | 12 |
| Local Shopping Sites | 0.14 | 0 | 2 |
| Local Travel/Tourism Sites | 0.05 | 0 | 1 |
| Local ISP Sites | 0.97 | 0 | 5 |
| Local Directory Sites | 0.01 | 0 | 1 |
| | | | |
| Number of Markets | 138 | | |

**Table 4: Is there More Local Online Content in Larger Markets?**

|  | (1) Total Locally Targeted Sites | (2) Local Web Service Provider Sites | (3) Local Commercial Online Network Sites | (4) Local Search Engine Sites | (5) Local Government Sites | (6) Local Education Sites | (7) Local Adult Sites |
|---|---|---|---|---|---|---|---|
| DMA Pop. 1990 (mil.) | 2.162 (0.087)** | 0.175 (0.028)** | 0.004 (0.005) | 0.027 (0.013)* | 0.079 (0.016)** | 0.539 (0.038)** | 0.349 (0.034)** |
| Constant | 2.577 (0.241)** | 0.191 (0.078)* | 0.008 (0.013) | 0.071 (0.036)* | 0.067 (0.044) | 0.109 (0.106) | 0.239 (0.093)* |
| Observations | 138 | 138 | 138 | 138 | 138 | 138 | 138 |
| R-squared | 0.82 | 0.22 | 0.01 | 0.03 | 0.15 | 0.59 | 0.44 |

|  | (8) Local Marketing/ Corporate Sites | (9) Local News/ Information/ Entertainment Sites | (10) Local Shopping Sites | (11) Local Travel/ Tourism Sites | (12) Local ISP Sites | (13) Local Directory Sites |
|---|---|---|---|---|---|---|
| DMA Pop. 1990 (mil.) | 0.171 (0.032)** | 0.558 (0.053)** | 0.034 (0.015)* | 0.008 (0.008) | 0.220 (0.034)** | -0.002 (0.003) |
| Constant | 0.298 (0.088)** | 0.846 (0.147)** | 0.090 (0.043)* | 0.038 (0.023) | 0.610 (0.095)** | 0.010 (0.009) |
| Observations | 138 | 138 | 138 | 138 | 138 | 138 |
| R-squared | 0.18 | 0.45 | 0.03 | 0.01 | 0.23 | 0.00 |

Standard errors in parentheses. * significant at 5% level; ** significant at 1% level

**Table 5: Does Local Content Induce People to Connect?**

|  | (1) | (2) | (3) | (4) | (5) | (6) | (7) | (8) |
|---|---|---|---|---|---|---|---|---|
|  | Net hm use comp/WTV own | Net hm use comp/WTV own | Local News/Info./ Entert. Sites | Net hm use comp/WTV own | Net hm use comp/WTV own | Net hm use comp/WTV own | Local News/Info./ Entert. Sites | Net hm use comp/WTV own |
|  | OLS | OLS | OLS | IV | OLS | OLS | OLS | IV |
| 1990 DMA Pop. (mil.) | -0.00024 (0.00104) | -0.00431 (0.00113)** | -0.95466 (0.46889)* | -0.00951 (0.00458)* | -0.00056 (0.00042) | -0.00182 (0.00092)* | -0.93696 (0.44332)* | -0.00490 (0.00323) |
| Local News/Info./Ent. Sites |  | 0.00827 (0.00201)** |  | 0.01885 (0.00845)* |  | 0.00258 (0.00160) |  | 0.00883 (0.00537) |
| College or Graduate School (mil.) |  |  | 8.62958 (2.61284)** |  |  |  | 8.54122 (2.47731)** |  |
| Controls | No | No | No | No | Yes | Yes | Yes | yes |
| Constant | 0.32077 (0.00820)** | 0.31135 (0.00806)** | 1.77587 (0.33182)** | 0.29930 (0.01179)** | 0.14589 (0.01232)** | 0.14299 (0.01264)** | 1.80556 (0.34422)** | 0.13596 (0.01411)** |
| Observations | 86523 | 86523 | 86523 | 86523 | 75311 | 75311 | 75311 | 75311 |
| R-squared | 0.00 | 0.00 | 0.78 |  | 0.23 | 0.23 | 0.78 | 0.23 |

Robust standard errors in parentheses. * significant at 5% level; ** significant at 1% level. Controls include race dummies, gender, income dummies, education dummies, age dummies, occupation and industry dummies, and variables reflecting household composition.

**Table 6: Does Racial Isolation Explain Connection?**

|  | (1) black | (2) nonblack | (3) MSA FE | (4) black | (5) nonblack | (6) MSA FE |
|---|---|---|---|---|---|---|
| MSA Percent Black | -0.103 (0.086) | -0.079 (0.057) |  |  |  |  |
| Black Dummy x MSA Pct. Black |  |  | -0.207 (0.097)* |  |  |  |
| Black Dummy |  |  | -0.087 (0.018)** |  |  | -0.115 (0.011)** |
| Black MSA Population |  |  |  | -0.013 (0.010) | 0.007 (0.010) |  |
| Non-Black MSA Population |  |  |  | 0.000 (0.002) | -0.004 (0.002) |  |
| Black Dummy x Black MSA Pop. |  |  |  |  |  | -0.034 (0.012)** |
| Black Dummy x Non-Black MSA Pop |  |  |  |  |  | 0.005 (0.002)* |
| Controls | yes | yes | yes | yes | yes | yes |
| Observations | 8757 | 65204 | 73961 | 8757 | 65204 | 73961 |
| R-squared | 0.19 | 0.21 | 0.24 | 0.19 | 0.21 | 0.24 |

Linear probability models. Dependent variable is whether individual has a home Internet connection. Robust standard errors in parentheses (clustering on MSA). * significant at 5% level; ** significant at 1% level. Controls include race dummies, gender, income dummies, education dummies, age dummies, occupation and industry dummies, and variables reflecting household composition.

**Table 7: Does Local Content Induce People to Own Computers?**

|  | (1) | (2) | (3) | (4) | (5) | (6) |
|---|---|---|---|---|---|---|
|  | One or More Computers at Home | One or More Computers at Home | One or More Computers at Home | One or More Computers at Home | One or More Computers at Home | One or More Computers at Home |
|  | OLS | OLS | IV | OLS | OLS | IV |
| DMA Pop. 1990 (mil.) | -0.00081 | -0.00534 | -0.00878 | -0.00109 | -0.00240 | -0.00366 |
|  | (0.00110) | (0.00158)** | (0.00424)* | (0.00076) | (0.00144) | (0.00275) |
| Local News/Information/Entertainment Sites |  | 0.00921 | 0.01621 |  | 0.00267 | 0.00524 |
|  |  | (0.00268)** | (0.00837) |  | (0.00222) | (0.00502) |
| Constant | 0.51497 | 0.50447 | 0.49650 | 0.26510 | 0.26209 | 0.25920 |
|  | (0.01090)** | (0.01077)** | (0.01424)** | (0.01500)** | (0.01537)** | (0.01479)** |
| Observations | 86523 | 86523 | 86523 | 75311 | 75311 | 75311 |
| Controls | No | No | No | Yes | Yes | Yes |
| R-squared | 0.00 | 0.00 | 0.00 | 0.31 | 0.31 | 0.31 |

Robust standard errors in parentheses. * significant at 5% level; ** significant at 1% level. Controls include race dummies, gender, income dummies, education dummies, age dummies, occupation and industry dummies, and variables reflecting household composition.

**Table 8: Computer Ownership, Market Size, and Racial Isolation**

|  | (1) | (2) | (3) | (4) | (5) | (6) |
|---|---|---|---|---|---|---|
|  | black | nonblack | All (MSA FE) | Black | nonblack | All (MSA FE) |
| MSA Black Pct. | -0.255 (0.117)* | -0.142 (0.072)* |  |  |  |  |
| Black Dummy x MSA Black Pct. |  |  | -0.282 (0.119)* |  |  |  |
| Black Dummy |  |  | -0.082 (0.021)** |  |  | -0.137 (0.014)** |
| Black MSA Population |  |  |  | -0.015 (0.021) | -0.004 (0.009) |  |
| Non-Black MSA Population |  |  |  | 0.003 (0.004) | -0.003 (0.002) |  |
| Black Dummy x Black MSA Pop. |  |  |  |  |  | -0.024 (0.015) |
| Black Dummy x Non-Black MSA Pop |  |  |  |  |  | 0.006 (0.003) |
| Controls | yes | yes | yes | Yes | yes | Yes |
| Observations | 8757 | 65204 | 73961 | 8757 | 65204 | 73961 |
| R-squared | 0.26 | 0.29 | 0.32 | 0.26 | 0.29 | 0.31 |

Linear probability models. Dependent variable is whether individual has a home computer. Robust standard errors in parentheses (clustering on MSA). * significant at 5% level; ** significant at 1% level. Controls include race dummies, gender, income dummies, education dummies, age dummies, occupation and industry dummies, and variables reflecting household composition.